\documentclass[
reprint,
superscriptaddress,
showpacs,
preprintnumbers,
amsmath,
amssymb,
aps, prx,
floatfix,
] {revtex4-2}

\usepackage{graphicx}  
\usepackage{dcolumn} 
\usepackage{bm}        
\usepackage{float}
\usepackage{siunitx}
\usepackage{xcolor}
\usepackage{comment}
\usepackage{bm}
\usepackage{amsthm}
\usepackage{ulem}
\usepackage{cancel}

\usepackage{graphicx}
\usepackage{upgreek}

\begin{document}

\author{Vedin Dewan}
\affiliation{Department of Mechanical and Aerospace Engineering, Princeton University, Princeton, New Jersey, 08544}

\author{Aleksei M.\ Zheltikov}
\email{zheltikov@tamu.edu}
\affiliation{Institute of Quantum Science and Engineering, Department of Physics and Astronomy, Texas A\&M University, College Station TX 77843}

\author{Julia M.\ Mikhailova}
\email{j.mikhailova@princeton.edu}
\affiliation{Department of Mechanical and Aerospace Engineering, Princeton University, Princeton, New Jersey 08544}

\title{Attosecond Nonlinear Quantum Electrodynamics in Laser-Driven Plasmas via Two-Photon Synchrotron Emission}

\begin{abstract}

Ultrafast strong-field laser--plasma physics is shown to offer a promising framework for relativistic nonlinear quantum electrodynamics (QED). As one of its key advantages, this approach to relativistic nonlinear QED does not require an external beam of relativistic particles. Instead, high-energy electrons are produced in this setting as a part of ultrafast strong-field laser--plasma interactions. An intense ultrashort laser pulse generates and accelerates dense electron bunches to relativistic energies, giving rise to photon-pair emission confined to the nanometer scale in space and the attosecond scale in time. As a lowest-order nonlinear QED process, relativistic electrons in laser-driven plasmas are shown to give rise to attosecond bursts of two-photon emission, providing an ultrabroadband source of correlated photon pairs. As a physically insightful estimate, the rate of this two-photon emission is expressed via a product $ \alpha^2 \gamma \omega_{turn}$, where $\alpha$ is the fine-structure constant, $\gamma$ is the Lorentz factor, and $ \omega_{turn}$ is the local relativistic curvature frequency. Photon pairs with strongest correlations, providing a resource for photon entanglement, are emitted at a much lower rate, estimated as $ \alpha^2 \gamma^2 \omega_{turn} E_{\perp} /E_S$, where $E_{\perp}$ is the laser electromagnetic field, determining the transverse Lorentz force, and $E_S$ is the Schwinger critical field. Our study offers a clear guidance on how quantum aspects of laser-driven relativistic plasma electrodynamics can be isolated from their classical counterparts, enabling a physically justifiable approach to the analysis of nonlinear QED phenomena in complex laser--plasma interactions driven by ultrashort high-intensity laser pulses. With purely quantum-field effects isolated via $\alpha^2 $  and $E_{\perp}/E_S$ factors, the role of classical plasma analysis in this framework is to simulate spatially and temporally resolved electron trajectories, thus providing direct in situ probes for local $\gamma$ and $ \omega_{turn}$ factors of plasma electrons. At laser intensities corresponding to relativistic laser amplitudes $a_0 \sim 20\text{--}200$, accessible at multi-petawatt laser facilities, laser--solid interactions drive electrons to $\gamma \sim 100\text{--}200$, and their correlated attosecond photon-pair emission emerges as a previously unexplored quantum channel competing with relativistic high-order harmonic generation.

\end{abstract}

\maketitle
Nonlinear relativistic quantum electrodynamics is a new frontier of laser science \cite{mourou2006optics, di2012extremely, mourou2019nobel}, offering a vast and promising area for the expansion of rapidly evolving laser technologies \cite{barty2016nexawatt, poder2018experimental, fedeli2021probing, danson2019petawatt, willingale2023status}. Even though electromagnetic fields attainable with the most advanced multipetawatt laser systems are many orders of magnitude lower than the Schwinger limit -- the strength of electromagnetic fields at which vacuum becomes polarizable, giving rise to electron–positron pairs, photon splitting, and light-by-light scattering \cite{berestetskii1982quantum}, a number of powerful concepts and approaches have been developed within the past years to make observation of laser-driven nonlinear QED effects possible \cite{mourou2006optics, di2012extremely, mourou2019nobel, gonoskov2022charged, elkina2011qed, nerush2011laser}.
Processes of multiphoton emission and nonlinear light scattering by relativistic particles are among the central themes in laser QED research \cite{bula1996observation, burke1997positron, li2015attosecond, edwards2016strongly, yan2017high, di2018implementing, mackenroth2013nonlinear, king2013}. Such processes not only serve as examples of genuine nonlinear QED phenomena, but may also enable scenarios for correlated-photon-pair and photon-entanglement generation. As a milestone along this path, theoretical analysis by Schützhold et al. \cite{schutzhold2006signatures, schutzhold2008tabletop} has shown that electrons moving in an extreme-intensity laser or undulator periodic field can drive vacuum fluctuations, giving rise to entangled photon pairs.  Analytical and numerical studies by Lötsted and Jentschura \cite{lotstedt2009correlated, lotstedt2009nonperturbative} have demonstrated the generation of entangled photon pairs via laser-driven double Compton backscattering. Recent numerical analysis by Zhang et al. \cite{zhang2023entangled} suggests that entangled x-ray photon pairs may be produced in self-amplification of spontaneous emission in free-electron lasers. 

Here, we explore an alternative approach to relativistic nonlinear QED, including, as the lowest-order nonlinear QED process, photon-pair and biphoton-entanglement generation. The nonlinear QED setting examined in this study builds upon nontrivial, often unexpected properties of relativistic plasmas driven by high-intensity ultrashort laser pulses. Relativistic plasma electrons in such a setting will be shown to give rise to attosecond bursts of two-photon emission, providing an ultrabroadband source of correlated photon pairs. Unlike the earlier proposals for nonlinear QED schemes for photon-pair generation, the approach that we analyze here does not require an externally generated beam of relativistic electrons. Instead, high-energy electrons are produced in our setting as a part of ultrafast strong-field laser--plasma interactions. 

To set the framework for our analysis, we consider interaction of a relativistic-intensity, few-cycle laser pulse with a solid-density plasma slab with a steep density profile. The laser driver, comprising $E_y$ and $B_z$ components, has a pulse width $\tau_L$ consisting of several laser field cycles $T_L$, where $T_L = 2\pi / \omega_0 = \lambda_0 / c$. The laser field strength $E_0$ ranges from a few to hundred times the relativistic field strength $E_{\text{rel}} = m_0c\omega_0/e$. Here, $\omega_0$ and $\lambda_0$ are the central frequency and wavelength of the laser driver, respectively; $c$ is the speed of light in vacuum; $m_0$ is the electron rest mass; and $e$ is the electron charge. The field strength is measured in the dimensionless relativistic laser amplitude units $a_0 = E_0/E_{rel} = eE_0/m_0c\omega_0$. The initial electron density $N_e$ in the plasma slab is typically set at the level of hundreds of critical densities $N_c=m{\omega_0}^2/(4\pi e^2)$ (in CGS units) for $\omega_0$. 

A characteristic picture of classical relativistic electrodynamics in such a laser--plasma interaction setting has been revealed by extensive particle-in-cell (PIC) simulations and understood in depth in terms of the key properties of radiation by accelerated charged particles \cite{Mikhailova2005generation, Mikhailova2012isolated, Edwards2016waveform, Edwards2020electron, Edwards2014enhanced, Edwards2016multipass, Edwards2020x, Brugge2010enhanced, Dromey2012coherent, Cherednychek2016analytical, Cousens2016temporal, fasano2023electron}. One of the most striking effects observed in this setting is that laser-driven electrons forming dense bunches \cite{fasano2023electron}, moving along sharply bent, synchrotron-like trajectories, coherently emit extraordinarily brief and intense bursts of extreme-ultraviolet to x-ray radiation in a specular direction, with pulse widths as short as a few attoseconds \cite{Mikhailova2005generation, Mikhailova2012isolated}. This coherent synchrotron emission (CSE) \cite{Mikhailova2012isolated, Edwards2016multipass, Cherednychek2016analyticala, Cousens2016temporal, Edwards2020x}  manifests in experiments and numerical simulations as relativistic high-order harmonic generation (HHG) \cite{Gibbon1996harmonic, Lichters1996short, Mikhailova2005generation, Tarasevitch2007transition, Dromey2009diffraction, Heissler2012few, Heissler2015multi, Heisler2010single, Teubner2009high, cavagna2024continuous, Dollar2013scaling, Vincenti2014optical}, and offers a promising route toward sources of high-energy photons and light intensities approaching the Schwinger limit \cite{Edwards2020x, vincenti2019achieving}.

\begin{figure} [hbp]
    	\centering
        \includegraphics[width=\linewidth]{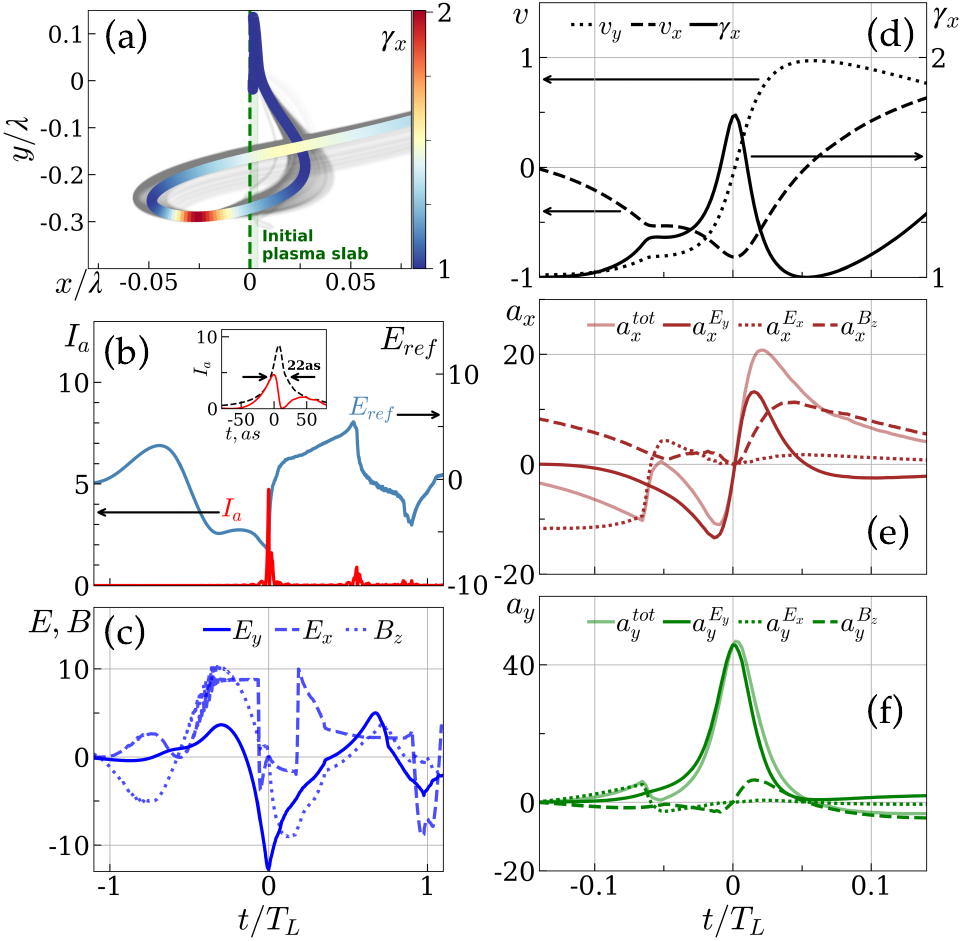}
    	\caption{Normal-incidence, weakly relativistic kinematics of leading electrons in the emitting bunch from PIC simulations of a plasma layer (thickness $D$ = 3.2nm, density $N_e= 500N_c$) driven by a laser pulse with strength $a_0 = 10$, FWHM duration $\tau_L$ = 3 fs, and incidence angle $\theta = 0^\circ$. (a) Representative synchrotron-like electron trajectories (grey lines) in the $x-y$ plane; the thick  multi-colored curve shows the average trajectory with color indicating the Lorentz factor $\gamma_x$ of electrons emitting in the specular direction ($x$). The initial plasma layer is marked by a vertical dotted line. (b) Time-resolved reflected electric field $E_{ref}$ (blue) and attosecond pulse intensity $I_{a}$ after spectral filtering in the range $10<\omega/\omega_0 <400$ (red). (c) Electric and magnetic fields seen by the moving electron. (d) Electron velocities and Lorentz factor $\gamma_x$ versus time. (e, f)Longitudinal ($a_x$) (e) and transverse ($a_y$) (f) accelerations highlighting the contributions of electric and magnetic field components to the corresponding Lorentz forces. All variables are in relativistic units.
     }
    	\label{fig:SynchrotronTrajectories}
  \end{figure}

Figure~\ref{fig:SynchrotronTrajectories} presents the results of PIC simulations performed with EPOCH \cite{EPOCH} for a plasma layer of thickness $D$ = 3.2 nm and electron density $N_e$ = 500$N_c$, driven by a laser pulse with central wavelength $\lambda_0$= 800 nm, field strength $a_0 = 10$ and FWHM pulse width $\tau_L$ = 3 fs, propagating along x-axis normally incident on the plasma layer. Figure~\ref{fig:SynchrotronTrajectories}(a) shows how leading electrons in the emitting bunch move along strongly bent trajectories, acquiring relativistic velocities along with high longitudinal Lorentz factor $\gamma_x$ (Fig.~\ref{fig:SynchrotronTrajectories}(d))and emitting attosecond bursts of extreme-ultraviolet radiation with intensity $I_a$ and pulse width of $\tau_{as} \approx$ 22as (Fig.~\ref{fig:SynchrotronTrajectories}(b)). All variables in this and other figures, including $I_a$, $E$, $B$, $v$, and $a$, are given in relativistic dimensionless units. The key feature of synchrotron-like emission is that the trajectory is strongly bent indicating a high transverse acceleration $a_y$ (Fig.~\ref{fig:SynchrotronTrajectories}(f)) near the time of emission perpendicular to a high longitudinal velocity $v_x$ (Fig.~\ref{fig:SynchrotronTrajectories}(d)). The acceleration of relativistic electrons is given by
\begin{equation}
\mathbf{a} = \frac{1}{m_0 \gamma} \left( \mathbf{F} - \frac{\mathbf{v}}{c^2} \left( \mathbf{v} \cdot \mathbf{F} \right) \right),
\end{equation}
where $\mathbf{F} = e \left( \mathbf{E} + \mathbf{v}\times \mathbf{B} / c \right)$ is the Lorentz force, the incident laser has electric field $E_y$ and magnetic field $B_z$, and the plasma space-charge electric field is $E_x$.
In the laser--plasma context, the role of the longitudinal quasi-electrostatic plasma space-charge force $E_x$ is to create conditions for the electron to move \textit{toward} the laser beam accelerated in tangential (specular reflection) direction by the longitudinal Lorentz force $F_{\parallel} = e\!\left(E_x + v_y B_z/c\right)$, as shown in Fig.~\ref{fig:SynchrotronTrajectories}(e). The turning of electron trajectory arises from transverse acceleration $a_y$, driven by the transverse Lorentz force $F_{\perp} = e\!\left(E_y - v_x B_z/c\right)$ around the peak of the longitudinal velocity $v_x$. The degree to which the emission of turning relativistic electron is frequency upshifted and beamed depends on the Lorentz factor $\gamma_x$. The characteristic synchrotron-like trajectories appear across a broad range of conditions in relativistic laser--solid interactions, including both normal and oblique laser incidence, in both thin and semi-infinite plasmas, and can acquire stronger curvature in the presence of a plasma density gradient \cite{Edwards2020x}.

Transverse acceleration $a_y$ and its individual components
\begin{equation}
\begin{split}
a_y^{\mathrm{tot}} = \frac{e}{m_0 \gamma}
\left[
E_y - \frac{v_x B_z}{c}
-\frac{v_y}{c^2} \left( \mathbf{v} \cdot \mathbf{E} \right)
\right]=\\
=\frac{e}{m_0 \gamma}
\left[
E_y - \frac{v_x B_z}{c}
- \frac{v_y}{ c^2} \left( v_y E_y + v_x E_x \right) 
\right]=\\
= a_y^{(E_y)} + a_y^{(B_z)} + a_y^{E_x},
\end{split}
\end{equation}
 are shown in Fig.~\ref{fig:SynchrotronTrajectories}(f). At the peak of emission, $t/T_L = 0$, the transverse acceleration $a_y$ and  trajectory curvature responsible for the synchrotron emission, are  driven by $F_{\perp}=e\!\left(E_y - v_x B_z/c\right)$, as $v_y=0$.

In a weakly relativistic interaction at $a_0 = 10$ in Fig.~\ref{fig:SynchrotronTrajectories}, the magnetic component of the transverse Lorentz force, $-ev_x B_z / c$, is small because $v_x \ll c$ , and the transverse acceleration $a_y$ is primarily driven by $E_y$.  At the emission point of the leading electron, the fields are $E_y = -12.03$, $B_z = 0.05$, $E_x = -0.01$, with $\gamma_x = 1.75$ (in relativistic units). For larger $a_0$, $v_x B_z / c$ becomes comparable to $E_y$ and can counteract or reinforce it, effectively decreasing or increasing the net turning acceleration $a_y$. The instantaneous angular turning frequency for the trajectory in Fig.~\ref{fig:SynchrotronTrajectories}(a) is given by
\begin{equation}
\omega_{\mathrm{turn}} =\frac{v_{\parallel}}{\rho}= \frac{a_y}{v_x}=\frac{e\left(E_{y} - v_{x} B_z/c\right)}{m_0 \gamma_{x} v_{x}}
\end{equation}
where $\rho=v^2_{\parallel}/a_{\perp}$ is the instantaneous radius of curvature of the electron trajectory at time of emission, governed by the transverse Lorentz force $e\!\left(E_y - v_x B_z/c\right)$ , $v_{\parallel}=v_x, a_{\perp}=a_y$, and $v_y=0$. 

\begin{figure} [hbp]
    	\centering
    	\includegraphics[width=\linewidth]{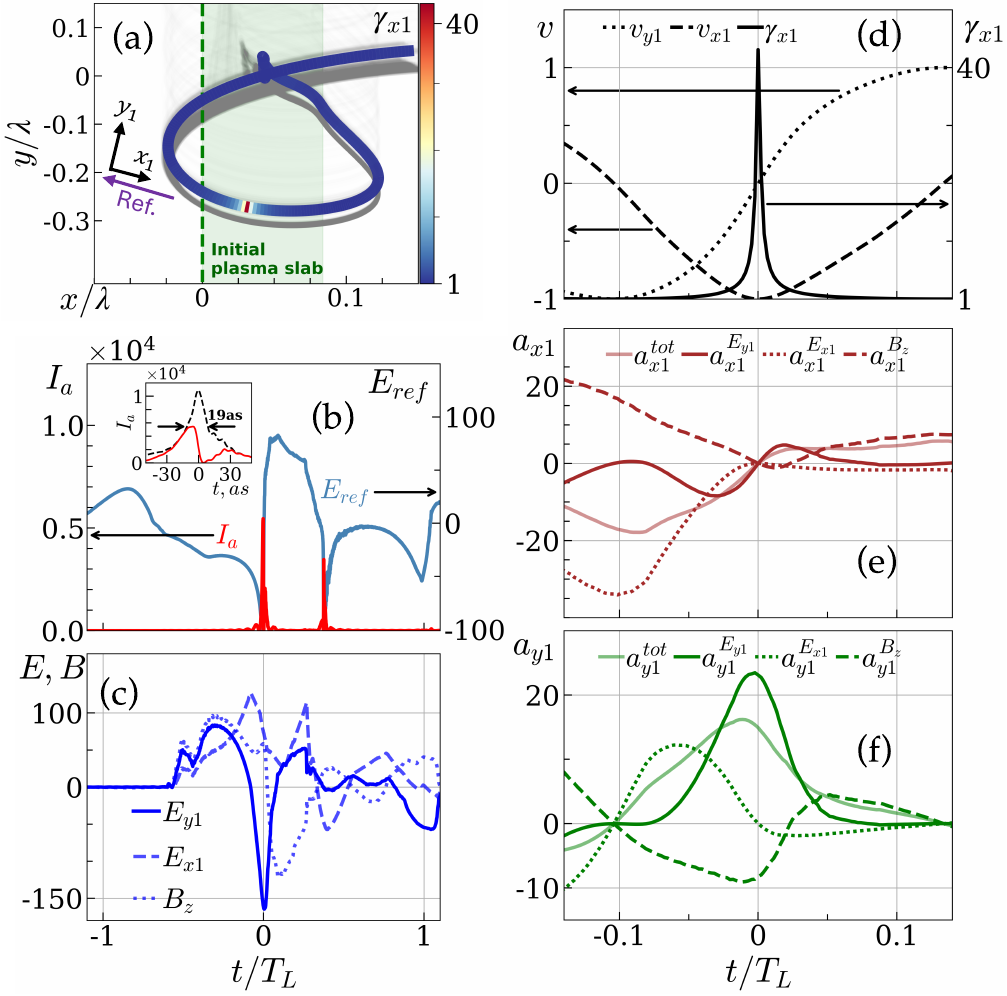}
    	\caption{Oblique-incidence, strongly relativistic kinematics of leading electrons in the emitting bunch from PIC simulations of a plasma layer (thickness $D$ = 67nm, density $N_e= 300N_c$) driven by a p-polarized laser pulse with strength $a_0 = 100$, FWHM duration $\tau_L$ = 3 fs, and incidence angle $\theta = 30^\circ$. (a) Representative synchrotron-like electron trajectories (grey lines) in the $x-y$ plane; the thick  multi-colored curve shows the average trajectory with color indicating the Lorentz factor $\gamma_x$ of electrons emitting in the specular direction ($x$). The initial plasma layer is marked by a vertical dotted line. (b) Time-resolved reflected electric field $E_{ref}$ (blue) and attosecond pulse intensity $I_{a}$ after spectral filtering in the range $10<\omega/\omega_0 <400$ (red). (c) Electric and magnetic fields seen by the moving electron. (d) Electron velocities and Lorentz factor $\gamma_x$ versus time. (e, f) Longitudinal ($a_x$) (e) and transverse ($a_y$) (f) accelerations highlighting the contributions of electric and magnetic field components to the corresponding Lorentz forces. All variables are in relativistic units.
     }
    	\label{fig:SynchrotronTrajectories30}
  \end{figure}
  
Figure~\ref{fig:SynchrotronTrajectories30} shows the strongly relativistic kinematics of emitting electrons for oblique incidence of a $p$-polarized laser pulse with $a_0=100$. Velocities, accelerations, and electric fields in Figs.~\ref{fig:SynchrotronTrajectories30} (c--f) are plotted in the rotated $(x_1, y_1)$ reference frame, where $x_1$ is aligned with the specular reflection direction, as shown in Fig.~\ref{fig:SynchrotronTrajectories30}(a). As shown in Fig.~\ref{fig:SynchrotronTrajectories30}(f), at  $a_0=100$, the magnetic component of the transverse Lorentz force, $v_{x1} B_z / c$, becomes comparable to $E_{y1}$ and counteracts it, effectively decreasing the net turning acceleration $a_{y1}$. At the emission point of the leading electron, the fields (in relativistic units) are $E_{y1} = -161.29$, $B_z = -57.42$, $E_{x1} = 73.89$, with $\gamma_{x1} = 43$. 

The classical synchrotron radiation in Figs.~\ref{fig:SynchrotronTrajectories}(b) and \ref{fig:SynchrotronTrajectories30}(b) is a result of continuous single-photon emission. Next, we consider the rate of correlated, discrete two-photon pair production arising from the same classical trajectories, but governed by quantum emission probability. To gain insights into the inner workings of nonlinear QED in strong-field ultrafast laser--matter interactions, we first seek to understand the fundamental space-time scales $\{\xi_i, \theta_j\}$ within which nonlinear QED phenomena tend to build up. By accomplishing this, we will be able to appreciate how the parameters of smallness in perturbative nonlinear QED naturally emerge from physically transparent combinations of these space--time scales $\{\xi_i, \theta_j\}$. Identifying the first two length scales in $\{\xi_i, \theta_j\}$ is straightforward. Indeed, the ratio of the classical electron radius, $r_e=e_0^2/(m_0 c^2 )\equiv \xi_1$ , to the Compton wavelength, $\lambda_C=\hbar /(m_0 c)\equiv \xi_2 $, $\xi_1 / \xi_2 = r_e / \lambda_C$ , is recognized as the fine-structure constant, $\alpha = e_0^2/(\hbar c)$. The significance of $\alpha$ as the smallness parameter in perturbative QED has been discussed in an extensive literature, with an illuminating overview found in the textbook \cite{berestetskii1982quantum}. 

The way the Compton wavelength $\lambda_C$ enters into the fine-structure constant $\alpha$ reflects its role as the amplitude of \textit{zitterbewegung} -- ultrafast quiver motion of relativistic electrons, which accompanies virtual transitions between positive- and negative-energy states of a Dirac particle with an emission of a virtual photon field. In its capacity as the amplitude of this quiver motion, $\xi_2=\lambda_C$ defines a measure of fundamental uncertainty in electron position. The ratio $\alpha=\xi_1/\xi_2 = r_e/\lambda_C$  is then better understood in the context of our laser--plasma interaction setting as the parameter that defines the extent to which the virtual photon field of a relativistic electron can couple out to the real photon field that this electron radiates as a part of its QED.

On the temporal side, one natural scale is the laser field cycle, $\theta_1=T_L$. For a Ti:Sapphire laser driver, i.e., radiation with $\lambda_0 \approx$ 800 nm, $\theta_1=T_L \approx$ 2.7 fs. While in classical nonlinear electrodynamics, $T_L$ is usually referred to as a benchmark for the shortest time of the problem, in quantum nonlinear electrodynamics, $\theta_1=T_L$ is, remarkably, the longest among the relevant time scales. The opposite extreme among $\theta_j$ is set by the \textit{zitterbewegung} cycle, $\theta_2=\hbar/(2m_0 c^2)$. Plugging the electron mass and electron charge into $\theta_2$ yields $\theta_2 \approx $ 4 zeptoseconds.

While the times $\theta_1$ and $\theta_2$ are universal, as they apply to any laser-driven nonlinear QED process, the QED buildup time, i.e., the time within which a laser field can couple to a relativistic electron, thus gaining an access to its virtual photon field, is QED-process-specific. For ultrafast strong-field laser--plasma interactions studied in this work, the classical relativistic electron trajectory determines the emission time window. Using the instantaneous relativistic turning rate, given by Eq (3), and taking into account that radiation is beamed into angle $\phi \sim 1/\gamma$, the emission duration can be estimated as the time required to traverse the beaming angle at the rate of electron rotation, given by
\begin{equation}
\theta_3 = \frac{\phi}{\omega_{\rm turn}}
= \frac{1}{\gamma \omega_{\rm turn}}
= \frac{m_0 v_{x}}{e\left(E_{y}-v_{x}B_z/c\right)}.
\end{equation}
It is within the time interval $\theta_3$ plasma electrons  pick up their acceleration (Figs. 1(d-f), 2(d-f)), acquiring signature, sharply bent sections of their trajectories (Figs. 1(a), 2(a)), thus being forced to emit attosecond bursts of synchrotron-type radiation (Figs. 1(b), 2(b) ).

\begin{figure}
    	\centering
    	\includegraphics[width=0.8\linewidth]{QED_Fig3.png}
    	\caption{Two-photon emission that accompanies electron transitions between stationary states with energy eigenvalues \( E_1 \) and \( E_2 \). This process gives rise to a pair of photons with frequencies \( \omega_1 \) and \( \omega_2 \), wave vectors \( \mathbf{k}_1 \) and \( \mathbf{k}_2 \), and polarization unit vectors \( \mathbf{e}_1 \) and \( \mathbf{e}_2 \).
     }
    	\label{fig:TwoPhotonDiagram}
  \end{figure}

With a minimal set of scales $\{\xi_i, \theta_j\}$ defined, we are in a position to start gaining insights into the smallness of perturbative, laser-driven relativistic nonlinear QED processes. Operating in the lowest order of nonlinearity, we consider a process whereby two photons with frequencies $\omega_1$ and $\omega_2$, wave vectors $\mathbf{k}_1$ and $\mathbf{k}_2$, and polarization unit vectors $\mathbf{e}_1$ and $\mathbf{e}_2$ -- $( \mathbf{k}_1, \omega_1,\mathbf{e}_1)$ and $( \mathbf{k}_2, \omega_2, \mathbf{e}_2)$, as shown via diagram in Fig.~\ref{fig:TwoPhotonDiagram} -- are emitted by a plasma electron as part of its relativistic QED, induced by a high-intensity ultrashort laser pulse. Because the probability for a single photon to be coupled out from the virtual photon field of an electron to the real radiation field scales with the fine-structure constant $\alpha$, the probability of emitting $n$ photons is proportional to $\alpha^n$.

 With all these factors in place, we find that the rate of two-photon emission in the QED process shown in Fig. ~\ref{fig:TwoPhotonDiagram} is proportional to a product of $\alpha^2$ and $\theta_3^{-1}$, leading to:
\begin{equation} \label{eq:1}
w \sim \frac{\alpha^2}{\theta_3} = \alpha^2 \gamma \omega_{turn}.
\end{equation}

The curvature of electron trajectories in Figs.~\ref{fig:SynchrotronTrajectories}(a) and ~\ref{fig:SynchrotronTrajectories30}(a) serves as a direct in situ probe for the gross local force acting on plasma electrons, suggesting $\overline{\omega}_{turn} = v_{\parallel}/\rho$ as a direct measure of the local relativistic cyclotron frequency $\omega_{turn}$, where $v_{\parallel}$ is the tangential electron velocity and $\rho$ is the curvature radius, which can be derived from numerical simulations such as those presented in Figs.~\ref{fig:SynchrotronTrajectories}(a) and ~\ref{fig:SynchrotronTrajectories30}(a).

Equation (\ref{eq:1}) has been derived without considering the coherence of photons in photon pairs. Therefore, the count rate $w$ in Eq.(\ref{eq:1}) includes both coherent photon pairs and pairs emitted via independent emission events in electron QED. To estimate a correction factor $\eta$ needed to isolate correlated photon pairs in the overall count rate $w$, defined by Eq.(\ref{eq:1}), we observe that photons $( \mathbf{k}_1, \omega_1,\mathbf{e}_1)$ and $( \mathbf{k}_2, \omega_2, \mathbf{e}_2)$ are coherent if they both are emitted from the same virtual photon field, i.e. the emission buildup scale for these photons is confined to $\xi_2 = \lambda_C$. The field needed to accelerate an electron at rest to the energy of $\sim m_0 c^2$ within $\xi_2 = \lambda_C$ is the Schwinger field $(E_S, B_S)$. Indeed, since $E_S = {m_0^2 c^3}/{(e_0 \hbar)}$, we find $e_0 E_S \xi_2 = m_0 c^2$. 

Therefore the correction factor $\eta$, quantifying the strength of quantum effects, is given by the ratio of the four-vector field $F_{\mu\nu}p^\nu$ in the electron's instantaneous rest frame to the Schwinger field:
\begin{equation}
    \eta = \frac{e\hbar}{m^3_0 c^4}\, \big|F_{\mu\nu} p^\nu\big|
=
\frac{\gamma}{E_S}
\left|\,\mathbf{E}_\perp + \frac{\mathbf{v}\times\mathbf{B}}{c}\,\right|.
\end{equation}
In our laser-plasma interaction scenario, for relativistic electrons moving along $x$ and accelerated transversely along $y$:
\begin{equation}
\eta =
\frac{\gamma_x \left( E_{y} - v_x B_z/c \right) }{E_S}.
\end{equation}

The count rate of coherent photon pairs is thus estimated as
\begin{equation} \label{eq:count}
w_c \approx \eta w \approx \alpha^2 \gamma \cdot \frac{E_{y} - v_x B_z/c}{E_S} \cdot \frac{e \left( E_{y} - v_x B_z/c \right)}{m_0v_x}. 
\end{equation}

Coherent photon pairs with count rate $w_c$ provide a resource of photon entanglement. Specifically, pairs of clockwise- and counterclockwise-polarized photons $( \mathbf{k}_1, \omega_1,\mathbf{e}_1 = \circlearrowright)$ and $(\mathbf{k}_2, \omega_2, \mathbf{e}_2 = \circlearrowleft)$, obeying the conservation  of angular momentum, combine into entangled states:
\[
\frac{1}{\sqrt{2}} \left( |\circlearrowright\rangle_A \otimes |\circlearrowleft\rangle_B + |\circlearrowleft\rangle_A \otimes |\circlearrowright\rangle_B \right),
\]

\noindent where $|\cdot\rangle_A$ and $|\cdot\rangle_B$ represent the quantum states of photons emitted along $\mathbf{k}_1$ and $\mathbf{k}_2$, respectively.

For a more rigorous QED treatment of strong-field laser--plasma interactions, we use the solution to the Dirac equation for electrons in an external electromagnetic field. From Figs. ~\ref{fig:SynchrotronTrajectories}(a, d) and ~\ref{fig:SynchrotronTrajectories30}(a, d) at the time of emission $t = 0$, we find $\rho \approx 0.015\lambda_0$ and $\rho \approx 0.07\lambda_0$ Lorentz factors $\gamma =1.73$ and $\gamma =43$ for the respective laser and plasma parameters. With $v_{\parallel}$ found from $\gamma$ via $v_{\parallel} = v_x = c \sqrt{1 - \gamma^{-2}}$, the typical buildup time of two-photon emission $\overline{\theta}_3 = (\rho/c) \left( 1 - \gamma^{-2} \right)^{-1/2}$, which simplifies to $\overline{\theta}_3 \approx (\rho/c) \left[ 1 + 1/(2\gamma^2) \right]$ for $\gamma \gg 1$, yields $\overline{\theta}_3 \approx 47$ as for $\gamma \approx 2$, and $\overline{\theta}_3 \approx 180$ as for $\gamma \approx 40$. Since this time is much shorter than the laser field cycle, $\theta_3 \ll \theta_1=T_L \approx 2.7$fs, the effective transverse field $E_\perp = E_y - v_x B_z/c$ can be considered constant within the entire two-photon emission buildup time. In the local-constant-field approximation, the Hamiltonian in the pertinent Dirac equation is then stationary \cite{SokolovTernov} in the instantaneous electron frame, allowing for stationary-state solutions for the electron wave functions $\psi(\mathbf{r}, t) = e^{\pm i E t / \hbar} \psi(\mathbf{r})$, where $E$ are the energy eigenvalues and $\psi(\mathbf{r})$ is the spatial part of the stationary-state wave function.
The energy eigenvalues of electron stationary states $\psi_{n s \kappa_{\parallel} \sigma}$ may then be written in analogy with the constant $H$-field synchrotron problem \cite{SokolovTernov} as
\begin{equation} \label{eq:3}
E=\hbar c K = \hbar c \sqrt{4\varepsilon_{\mathrm{eff}} n + \kappa_{\parallel}^2+\kappa_0^2},
\end{equation}
\noindent where $n$ and $s$ are the principal and radial quantum numbers, $\hbar\kappa_\parallel$ is the electron momentum component along the instantaneous direction of motion, $\kappa_0=m_0c/\hbar$,  $\varepsilon_{\mathrm{eff}}=
{e_0 |E_\perp|}/{2\hbar c}={e_0}/{2\hbar c}\left|E_y-{v_x B_z}/{c}\right|$.

The probability of correlated two-photon emission is found in this framework  via a perturbative treatment of two-photon transitions:
\begin{equation} \label{eq:4}
\begin{split}
(e^-; n_1, s_1, \kappa_{\parallel 1}, \sigma_1) \rightarrow (e^-; n_2, s_2, \kappa_{\parallel 2}, \sigma_2) + \\
(\gamma_1; \omega_1, \mathbf{k}_1, \mathbf{e}_1) + (\gamma_2; \omega_2, \mathbf{k}_2, \mathbf{e}_2), 
\end{split}
\end{equation}

\noindent between electron states $\psi_{n_1 s_1 \kappa_{\parallel 1} \sigma_1}$ and $\psi_{n_2 s_2 \kappa_{\parallel 2} \sigma_2}$, accompanied by the emission of two photons with frequencies $\omega_1$ and $\omega_2$, wave vectors $\mathbf{k}_1$ and $\mathbf{k}_2$, and polarization vectors $\mathbf{e}_1$ and $\mathbf{e}_2$ (Fig.~\ref{fig:TwoPhotonDiagram}), where the transition is driven by the curvature of the classical electron trajectory in the locally constant effective field. Various aspects of such a treatment have been developed in earlier work \cite{ritus1972vacuum, morozov1975elastic, sokolov1976two}. Specifically,  closed-form analytical solutions for the matrix elements for similar transitions in constant magnetic field have been presented by Sokolov et al. \cite{sokolov1976two}.

Expressing the pertinent conservation laws are the products $\delta(K_2 - K_1 + k_1 + k_2)\;\delta\!\left(\kappa_{\parallel 2} - \kappa_{\parallel 1} + k_{1\parallel} + k_{2\parallel}\right)$ emerging as isolated factors in these solutions. Photons emitted in the same two-photon emission event (\ref{eq:4}) are thus bound to have frequencies that satisfy energy conservation:
\begin{equation}  \label{eq:5}
E_1 = E_2 + \hbar \omega_1 + \hbar \omega_2.
\end{equation}
Momentum conservation is imposed along the instantaneous longitudinal direction,
\begin{equation}  \label{eq:6}
\kappa_{\parallel 1}
=
\kappa_{\parallel 2}+k_{1\parallel}+k_{2\parallel},
\end{equation}
where $k_{i\parallel}=\mathbf{k}_i\cdot\hat{\mathbf{v}}$ and $\hat{\mathbf{v}}$ is the unit vector along the instantaneous electron velocity:
\begin{equation}
\kappa_{\parallel 1} =\kappa_{\parallel 2} + k_1 \cos\phi_1 + k_2 \cos\phi_2,
\end{equation} 
\noindent where $\phi_1$ and $\phi_2$ are the angles between the emitted photon wave vectors $\mathbf{k}_1$ and $\mathbf{k}_2$ and the electron velocity.

Specifically, for photons of equal frequencies $\omega_1 = \omega_2 = \omega$, emitted symmetrically with respect to the instantaneous electron velocity, $(\phi_1 = \phi_2 = \phi)$, Eqs.(\ref{eq:5}) and (\ref{eq:6}) dictate the following relation between the frequency $\omega$ and the emission angle $\phi$:
\begin{equation}
\omega= \frac{E_1 q}{2\hbar \sin^2\phi}
\left[1-\sqrt{1-\frac{4\varepsilon_{\mathrm{eff}}\Delta n}{q K_1^2}\sin^2\phi}\right],
\end{equation}
where
\begin{equation}
q = 1-\left(\frac{\kappa_{\parallel 1}}{K_1}\right)\cos\phi .
\end{equation}

Under conditions where  $|E_\perp|/E_S \ll 1$, ${E_{1,2}}/{m_0 c^2} \gg 1$, and over a time interval where the field is locally constant, the probability of two-photon emission (\ref{eq:4}) per unit angle per unit time is (similarly to \cite{sokolov1976two}):
\begin{equation}  \label{eq:8}
w
=
\frac{25}{24}\alpha^2
\left(
\frac{e_0 |E_\perp|}{m_0 c}
\right)
\left(
\frac{E_1}{m_0 c^2}
\right)
\left(
1-\frac{3\chi}{5}-\frac{4\chi}{\sqrt{3}}
\right),
\end{equation}
where the invariant quantum parameter is
\begin{equation}
\chi
=\frac{E_1}{m_0 c^2} \cdot \frac{|E_\perp|}{E_S}
=\frac{E_1}{m_0 c^2} \cdot \frac{\left|E_y-v_x B_z/c\right|}{E_S}.
\end{equation}

With ${E_1}/{m_0c^2}$ in Eq.(\ref{eq:8}) recognized as $\sim \gamma$ factor, the leading term in Eq.(\ref{eq:8}) recovers the two-photon count rate $w$ in Eq.(\ref{eq:1}) up to a factor of $25/24$. Moreover, the first correction to the leading term in Eq.(\ref{eq:8}), i.e., the $\sim \chi$ terms in parentheses in Eq.(\ref{eq:8}) is fully consistent with Eq.(\ref{eq:count}). The parameter $\chi$, defining the order of smallness of this correction, is of the same order of smallness as the parameter $\eta$ in our much less rigorous, but, perhaps, more physically intuitive earlier treatment (cf. Eq.(\ref{eq:count}) and Eq.(\ref{eq:8})).

With Eqs. (\ref{eq:1}) and (\ref{eq:count}) providing a clear view of the physics behind the mathematical structure of the two-photon emission rate inherent in laser-driven relativistic plasma electrons [Eq.(\ref{eq:8})], we can make important conclusions regarding the methodological framework for the treatment of nonlinear QED phenomena in laser--plasma interactions driven by ultrashort high-intensity laser pulses. Laser-driven relativistic plasma electrodynamics in such a setting is, of course, too complex to leave any hope that a fully consistent quantum treatment of this dynamics could ever be productive. Eqs.(\ref{eq:1}), (\ref{eq:count}) and (\ref{eq:8}) are very helpful in this regard as they provide a clear guidance on how quantum aspects of laser-driven relativistic plasma electrodynamics can be isolated and separated from their classical counterparts, enabling physically justifiable approaches to the analysis of nonlinear QED effects in such settings.
Specifically, as one of their most powerful insights, Eqs.(\ref{eq:1}) and (\ref{eq:8}) clearly show that the two-photon emission rate of relativistic electrons can be represented as a product of three physically transparent factors -- the fine-structure constant $\alpha$, the Lorentz factor $\gamma$, and the local turning frequency $\omega_{turn}$. In this product, the fine structure constant $\alpha$ is responsible for quantum aspects of relativistic electrodynamics. As articulated above, the $\alpha^2$ factor in Eqs.(\ref{eq:1}) and (\ref{eq:8}) defines the extent to which the virtual photon field of a relativistic electron can couple out to the real two-photon field that this electron radiates as a part of its QED. The $\gamma$ and $\omega_{turn}$ factors, on the other hand, can be viewed as quantifiers of classical properties of relativistic plasma electrons and can be found by means of classical plasma electrodynamics, e.g., via PIC simulations  \cite{Mikhailova2012isolated, Edwards2016waveform, Edwards2020electron, Edwards2014enhanced, Edwards2016multipass, Edwards2020x, Brugge2010enhanced, Dromey2012coherent, Cherednychek2016analytical}.

Such a separation between the classical and quantum aspects of plasma electrodynamics is fully equivalent, as comparison of Eqs.(\ref{eq:1}) and (\ref{eq:8}) shows, to a framework in which the high-intensity laser driver field is treated classically, relativistic electrons in an external field are described in terms of Dirac wave functions, and the two-photon radiation field is viewed as a quantum field. The role of classical plasma analysis in this framework is to define the effective local $\gamma$ and $\omega_{turn}$ factors for plasma electrons. PIC simulations are ideally suited for this task, as they provide powerful means to resolve representative electron trajectories, such as those shown in Figs. \ref{fig:SynchrotronTrajectories} and \ref{fig:SynchrotronTrajectories30}), where velocities are local measures of $\gamma$ and the space-resolved curvature is a direct \textit{in situ} probe for the local $\omega_{turn}$ via $\overline{\omega}_{turn} = v_{\parallel}/\rho$.
Within the realm of perturbative nonlinear QED, photon pairs with strong quantum correlations are isolated, as Eqs.(\ref{eq:1}) and (\ref{eq:8}) show, by multiplying the total two-photon emission output by $\eta \approx \chi \sim \gamma E_{\perp}/E_S$. The $\eta \approx \chi$ multiplier is the smallness parameter of perturbative nonlinear QED, purely quantum in nature. 

To illustrate this approach, we resort once again to PIC simulations in Figs. \ref{fig:SynchrotronTrajectories} and \ref{fig:SynchrotronTrajectories30} to read out $v_{\parallel}$ and $\rho$ from electron trajectories around time of emission $t \approx 0$, we find 
\begin{equation*}  \label{eq:9}
\begin{split}
\overline{\omega}_{turn} = c/\rho \left[ 1 - \gamma^{-2})\right]^{1/2} \approx c/\rho \left[ 1 - 1/(2\gamma^2)\right] \\
\approx 2.12 \times 10^{16} \, \text{s}^{-1},
\end{split}
\end{equation*}
\noindent
for $\gamma = 1.73$ in Figs.\ref{fig:SynchrotronTrajectories}(a,d),
\begin{equation*}  
\overline{\omega}_{turn} \approx c/\rho \left[ 1 - 1/(2\gamma^2)\right] \approx 5.7 \times 10^{15} \, \text{s}^{-1},
\end{equation*}
\noindent
for $\gamma=43$ in Figs.\ref{fig:SynchrotronTrajectories30} (a,d). 

The total two-photon count rates are, respectively, 
\begin{equation*} 
w \approx \alpha^2 \gamma \cdot \overline{\omega}_{turn} \approx  1.95 \times 10^{12} \, \text{s}^{-1} 
\end{equation*}
for $\gamma=1.73$ in Fig. \ref{fig:SynchrotronTrajectories}, and
\begin{equation*} 
w \approx \alpha^2 \gamma \cdot \overline{\omega}_{turn} \approx 1.3 \times 10^{13} \, \text{s}^{-1} 
\end{equation*}
for $\gamma=43$ in Fig. \ref{fig:SynchrotronTrajectories30}.
This yields approximately $n \approx w \tau_{as} \approx 43 \times 10^{-6}$ photon pairs per pulse with the attosecond-scale pulse duration $ \tau_{as} \approx 22$as in Fig. \ref{fig:SynchrotronTrajectories}, and $n \approx 2.5 \times 10^{-4}$ photon pairs per pulse with $\tau_{as} \approx 19$as in Fig. \ref{fig:SynchrotronTrajectories30}. With the electron density $N_e \approx 500 N_c \approx 8.5 \times 10^{23}$ cm$^{-3}$, as in the simulations shown in Fig. \ref{fig:SynchrotronTrajectories}, and $N_e \approx 300 N_c \approx 5 \times 10^{23}$cm$^{-3}$, as in Fig. \ref{fig:SynchrotronTrajectories30} and considering a laser-irradiated plasma volume of $V \approx \lambda_0^3$, the two-photon count rate yields $ N = n N_e V \approx 2 \times 10^7 $ photon pairs per laser pulse for Fig. \ref{fig:SynchrotronTrajectories} and $N \approx 6 \times 10^7$ photon pairs per laser pulse for Figs. \ref{fig:SynchrotronTrajectories}(d--f). Assuming a repetition rate of laser pulses $f \approx 1$ kHz, these numbers translate into count rates of $Nf \approx 2 \times 10^{10}$ photon pairs per second and $Nf \approx 6 \times 10^{10}$ photon pairs per second, respectively.
With the laser driver field set at $10 E_{\text{rel}}$, as in the simulations shown in Fig. \ref{fig:SynchrotronTrajectories}, the smallness parameter $\eta = (\gamma E_{\perp}) / E_S$ is approximately $\eta \approx 6 \times 10^{-5}$ for electrons with $\gamma \approx 2$. 
With the laser driver field set at $100 E_{\text{rel}}$, as in the simulations shown in Figs. \ref{fig:SynchrotronTrajectories30}, the smallness parameter $\eta = (\gamma E_{\perp})/ E_S$ is approximately $\eta\approx 10^{-2}$ for electrons with $\gamma \approx 43$. The total number of correlated photon pairs is then $N_{coh} \approx 1287$ photon pairs per laser pulse for the electrons in Fig.\ref{fig:SynchrotronTrajectories}, and $N_{coh} \approx 9.5\times10^5$ photon pairs per laser pulse for those in Fig.\ref{fig:SynchrotronTrajectories30}. This leads to count rates of $N_{coh} f \approx 1.3 \times 10^6$ photon pairs per second and $N_{coh}  f \approx 9.5 \times 10^8$ photon pairs per second, respectively, assuming a laser and attosecond pulse repetition rate of $ f \approx 1 $kHz.

\begin{figure} [ht]
    	\centering
    	\includegraphics[width=\linewidth]{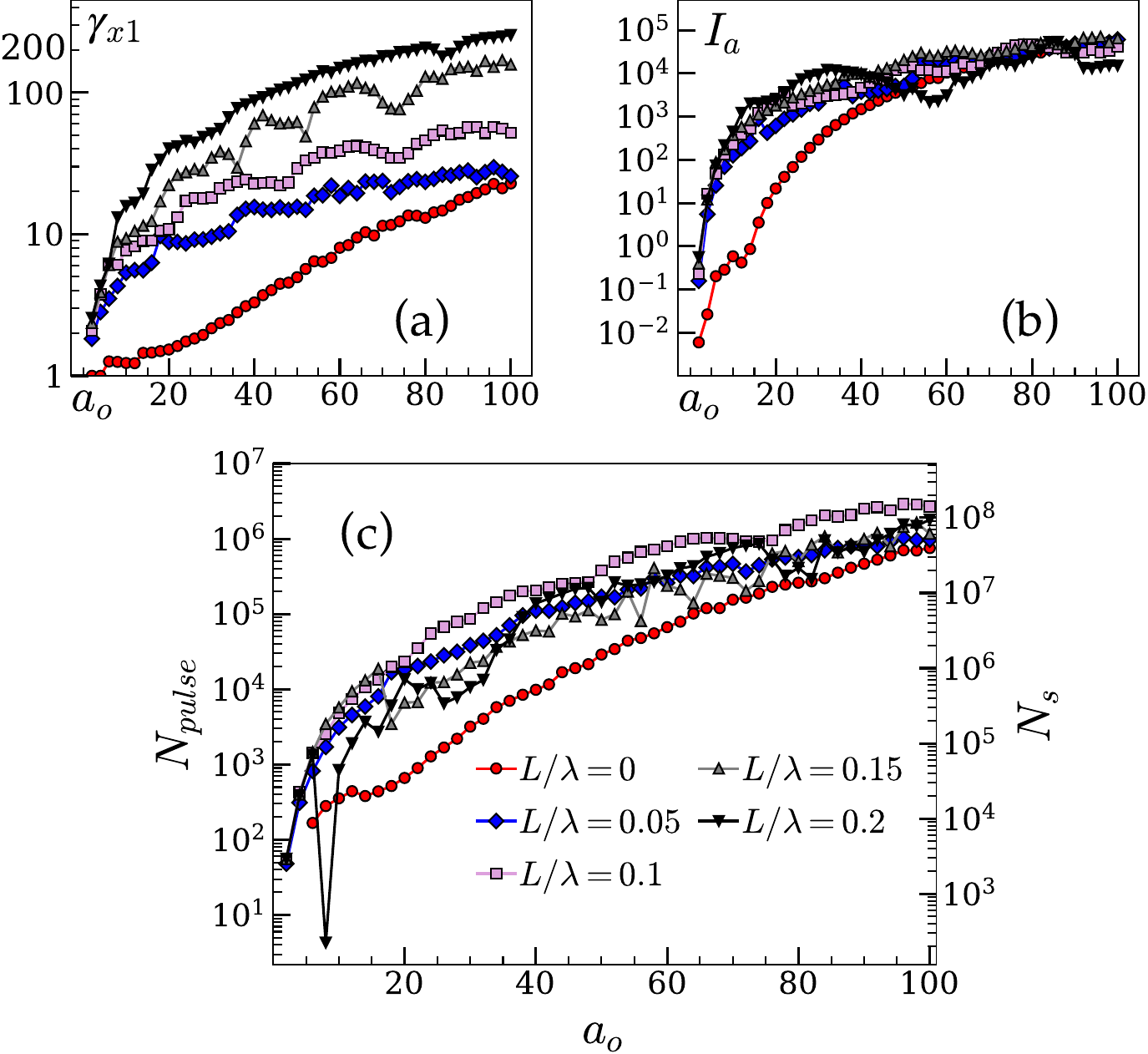}
    	\caption{
Scaling of electron energy, attosecond emission, and correlated photon-pair yield with $a_0$ and density gradient. (a) Peak Lorentz factor in the emission direction $\gamma_{x1}$; (b) Attosecond pulse intensity $I_a$ in relativistic units; (c) Number of correlated photon pairs per attosecond pulse $N_{pulse}$ and per second $N_s$ for parameters of the ALEPH~\cite{Wang2017ALEPH} petawatt system ($30fs$ pulses at repetition rate $f =$ 3.3Hz) versus driving laser field $a_0$ for $N_e = 300 N_c$, $D = \lambda_0$, $\theta = 45^\circ$, and five different density gradient scale lengths $L/ \lambda_0$ = 0, 0.05, 0.1, 0.15, and 0.2. The attosecond pulse intensity (in relativistic units) is obtained by filtering harmonics in the range $\omega/\omega_0 \in [10,400]$. Despite significantly higher $\gamma_{x1}$ values at larger gradients, the attosecond pulse intensity $I_a$ asymptotes to the same value near the optimal $a_0$, corresponding to $a_0/N \approx 0.33$. This suggests that, at optimal $a_0$, larger gradients can lead to a higher fraction of emitted photons exhibiting non-classical QED behavior.}
    	\label{fig:GammaPIC}
  \end{figure}

As a guide to understand the overall behavior of photon-pair production as a function of the laser intensity and plasma density, Fig. \ref{fig:GammaPIC} presents the dependence of the $\gamma$ ratio on the driving laser field $a_0$. Each data point in Fig. \ref{fig:GammaPIC} represents a single PIC simulation run, with $a_0$ ranging from 2 to 100, $N = 300N_c, D = \lambda_0, \theta = 45^\circ$. In earlier work \cite{Gordienko2005scalings,Edwards2020x} the ratio $a_0 / N$ has been identified as a similarity parameter, governing the efficiency of attosecond pulse generation in relativistic laser–plasma interactions, with $a_0/N \sim 0.33$ shown to be optimal for oblique incidence at around $\theta = 30 - 45^\circ$ \cite{Edwards2020x}. Figure. \ref{fig:GammaPIC} show that while larger density gradients lead to significantly higher electron Lorentz factors, the intensity of the classical attosecond pulse asymptotes to a value obtained in the optimal regime with $a_0/N \approx 0.33$. This behavior indicate that the additional electron energy at larger gradients may redistribute emission into quantum two-photon process, suggesting that a larger fraction of the emitted radiation may be carried by correlated photon pairs. Consequently, the relative contribution of nonclassical emission is expected to increase at larger gradients and larger $a_0$.

For longer, 30-fs driving pulses, the sub–laser-cycle electron trajectories and the resulting attosecond two-photon emission repeat al least once per laser cycle within the pulse envelope. This would lead to an increase in correlated photon-pair yield by about an order of magnitude compared to a single-cycle, 3-fs driver. Therefore, with petawatt systems such as the ZEUS \cite{maksimchuk2025zeus}, ALEPH \cite {Wang2017ALEPH}, and Extreme Light Infrastructure \cite{Gales2018}, operating with $\sim30$-fs pulses at $a_0 \sim 20 - 200$ and a repetition rate of $f \approx 1-5$ Hz in burst mode, the count rate of correlated photon pairs for $\gamma=100$, $\eta\sim 10^{-2}$, and $\tau_{as}\approx20$as can reach $N_{MPW} \approx \eta \alpha^2 \gamma \overline{\omega}_{turn} \cdot \tau_{as} \cdot \tau_L/T_L \cdot N_eV \cdot f \approx 3 \times 10^{8}$ photon pairs per second, corresponding to a substantial and experimentally accessible photon-pair flux, even after accounting for realistic detection efficiencies of extreme-ultraviolet radiation.

To summarize, we have shown that ultrafast strong-field laser--plasma physics offers a promising framework for relativistic nonlinear QED. As one of its key advantages, this approach to relativistic nonlinear QED does not require an externally generated beam of relativistic electrons. Instead, high-energy electrons are produced in this setting as a part of ultrafast strong-field laser--plasma interactions. A strong-field ultrashort laser pulse accelerates these electrons to relativistic energies, giving rise to photon-pair emission confined to the nanoscale in space and the attosecond scale in time. As a lowest-order nonlinear QED process, relativistic electrons in laser-driven plasmas give rise to attosecond bursts of two-photon emission, providing an ultrabroadband source of correlated photon pairs. As a physically insightful estimate, the rate of this two-photon emission is expressed via a product $\alpha^2\gamma\omega_{turn}$. Photon pairs with the strongest correlations, providing a resource for photon entanglement, are emitted at a much lower rate, estimated as $\alpha^2 \gamma^2 \omega_{turn} {E_{\perp}}/{E_S}$. 
Our study offers a clear guidance on how quantum aspects of laser-driven relativistic plasma electrodynamics can be isolated from their classical counterparts, enabling a physically justifiable approach to the analysis of nonlinear QED phenomena in complex laser--plasma interactions driven by ultrashort high-intensity laser pulses. With purely quantum-field effects isolated via $\alpha^2$ and ${E_{\perp}}/{E_S}$ factors, the role of classical plasma analysis in this framework is to simulate spatially and temporally resolved electron trajectories, thus providing direct \textit{in situ} probes for local $\gamma$ and $\omega_{turn}$.
For $a_0 \sim 20 - 200$, achievable at multi-petawatt laser user facilities such as the ZEUS \cite{maksimchuk2025zeus}, ALEPH \cite {Wang2017ALEPH}, ELI \cite{Gales2018}, and other facilities \cite{Pirozhkov2017,  Guo2018,  Bromage2019, Yoon2021}, electrons in laser--solid interactions can reach Lorentz factors $\gamma \sim 100 - 200$, entering a strongly relativistic regime. In this regime, correlated attosecond photon pair emission becomes significant and can compete with classical relativistic high-order harmonic generation, thus becoming a new quantum emission channel in relativistic laser--plasma interactions.

\begin{acknowledgments}
This work was supported by the Gordon and Betty Moore Foundation under Grant No. GBMF12255, DOI 10.37807/gbmf12255, by the National Science Foundation under Grant No. PHY 2512131 and PHY 2206711, and by the Department of Energy under Grant No. DE-SC0025497. V.D. gratefully acknowledges support from the Program in Plasma Science and Technology of Princeton University.
\end{acknowledgments}

\bibliography{References}

\end{document}